\documentclass[12pt]{iopart}
\usepackage{graphicx}

\begin{document}

\title[Possible GeV counterpart associated with Fermi LAT gamma-ray bursts]{Possible GeV counterpart at the ground level associated with Fermi LAT gamma-ray bursts}

\author{C. R. A. Augusto, C. E. Navia, M. N. de Oliveira}
\address{Instituto de F\'{\i}sica, Universidade Federal Fluminense, 24210-346,
Niter\'{o}i, RJ, Brazil} 

\author{Andr\'{e} Nepomuceno}
\address{Departamento de Ci\^{e}ncias da Natureza, IHS, Universidade Federal Fluminense, 28890-000, Rio das Ostras, RJ, Brazil}


\author{V. Kopenkin}
\address{ Research Institute for Science and Engineering, Waseda University, Shinjuku, Tokyo 169, Japan.}

\author{T. Sinzi}
\address{Rikkyo University, Toshima-ku, Tokyo 171, Japan}

\ead{navia@if.uff.br}
\vspace{10pt}
\begin{indented}
\item[]July 2017
\end{indented}

\begin{abstract}
From June 2014 to February 2017, the Fermi LAT detected 46 gamma-ray bursts (GRBs) with photon energies above 20 MeV, and the trigger coordinates of seven of them were within the FoV of New-Tupi detector located in the central region of the South Atlantic Anomaly (SAA).We show in this paper that two of these seven GRBs have a probable GeV counterpart observed at ground level by New-Tupi detector.
The first is GRB 160609A, a short duration GRB with a bright emission of photons over a broad energy range extending up to GeV energies. The second is GRB 160625B, a very long duration GRB, for which the Fermi LAT detected more than 300 photons with energies above 100 MeV in the $\sim$ 1 ks interval after the GBM trigger.
In the first case, the signal at New-Tupi has a nominal significance of $3.5\sigma$ in the counting rate time profiles, within the  T90($=5.6$ s) duration on Fermi GBM. However, the
effective significance is only $3.0\sigma$. In the second case, New-Tupi detector registered at least two excess (peaks) with a nominal statistical significance of $4.8\sigma$ and $5.0\sigma$
at 438 s and 558 s after the trigger. The first is within the $T90(=460$ s) on Fermi GBM.
Even so, the effective significance is only $\sim 2.0\sigma$.
In addition, from a Monte Carlo analysis, we show that the expected signal-to-noise ratio is compatible with the observation of GRB 160709A, only if the differential index of the GRB energy spectrum be equal or higher than -2.2 (a non-steep spectrum).
\end{abstract}

%
%
%
%
%

\section{Introduction}
\label{sec:intro}

The Fermi Large Area Telescope (Fermi LAT)  is the principal scientific instrument on the Fermi Gamma-Ray Space Telescope (Fermi) mission, covering the energy range from 20 MeV to more than 300 GeV. 
Another science instrument on the payload is the Fermi Gamma-ray Burst Monitor (Fermi GBM), which provides a good overlap with the LAT making observations in the range $\sim 8$ keV to $\sim 40$ MeV.
The Fermi GBM detects $\sim 250$ gamma-ray bursts (GRBs) per year. 
About half of them are located within the Fermi LAT FoV \cite{acke13}. 
However, only $\sim 10\%$ of GRBs were detected by the Fermi LAT ($\geq$ 100 MeV). 
The Fermi LAT continues successful exploratory gamma ray astronomy missions performed by the Energetic Gamma Ray Experiment Telescope (EGRET) on-board the Compton Gamma Ray Observatory (CGRO) \cite{sommer94,dingos94,schneid92,hurley94}.
Besides observations of GRBs in the MeV-GeV energy band, Fermi LAT has observed gamma ray emissions from the Galactic center \cite{abdo09a,abdo09c}, gamma ray emissions from supernova remnants, blazars and pulsars \cite{abdo09b}, 
including othrt gamma-rays emitters such as radio galaxies, starbust galaxies, binary systems, novae, diffuse emission,
and solar flares \cite{acke13}

Several scenarios have been suggested to explain a possible high energy (GeV band) component of GRBs, such as the synchrotron self-Compton (SSC) model \cite{panaitescu00,kumar08} that can explain the optical and gamma-ray correlation seen in some of GRBs data. 
It also implies that a relatively strong second order inverse Compton (IC) component of the GRB spectrum should peak in the energy range of tens GeV \cite{racusin08}. 

On the other hand, there are various efforts to observe at ground level GeV-TeV counterparts of GRBs.
Typically, the GRBs are detected by a statistical excess in the counting rate over the background noise.
Some of the experiments are highlighted here.

Milagro was a wide field (2 sr) high-duty cycle ($> 90\%$) ground-based water Cherenkov detector (60 m wide × 80 m long × 8 m deep) covered with a light-tight cover located at 2630 m above sea level in the Jemez Mountains, New Mexico. 
Milagro operated from January 2000 to May 2008.

As reported by Milagro, none of the events in the 2000-2006 sample showed significant very high-energy (VHE) emission \cite{abdo07a}. 
Even so, Milagro succeeded in the detection of TeV gamma rays from the galactic plane \cite{abdo07b}, and the  TeV gamma-ray emission from the Cygnus region of the Galaxy \cite{abdo07c}. 

The ARGO-YBJ detector at YangBaJing, China (Tibet $\sim 4300$ m a.s.l.) has a large active surface of around $6700 m^2$ of Resistive Plate Chambers, a wide field of view $\sim 2$ sr,  and a high duty cycle ($>$ 86 \%) \cite{bart14}.
ARGO-YBJ performed the search for GRB  emission in the energy range from 1 to 100 GeV in coincidence with satellite detection. 
From 17 December 2004 to 7 February 2013 a total of 206 GRBs occurring within the ARGO-YBJ field of view (zenith angle $θ \leq 45^{0}$) were investigated searching for an increase in the detector counting rates, and no significant excess has been found.

The HAWC Gamma-ray Observatory is a wide field of view ( $\sim 2$ sr) very high energy (VHE) gamma-ray extensive air shower (EAS).
HAWC is located at an altitude of 4100 m above sea level in Mexico (Sierra Negra).
With an array of 300 water Cherenkov detectors, HAWC  has an order of magnitude better sensitivity than its predecessor, the Milagro experiment.
The HAWC experiment can observe the prompt GRB phase and probe a spectrum at TeV energies.
The reported estimated rate may be as high as 1–2 GRBs per year.
So far, no statistically significant excess of counts has been found and upper limits have been estimated \cite{abey15}.

Many other experiments have searched for the GeV-TeV counterparts of GRBs, such as INCA \cite{vern00}, Tibet AS \cite{amer96}, HEGRA AIROBICC \cite{padi98}, GRAND \cite{poir03}, LAGO \cite{alla08}, the Cherenkov detector MAGIC \cite{albe06}. 
Essentially there were no significant TeV counterparts associated with the GRBs observed by satellites.

In the past three years (since 8 June 2014) New-Tupi detector situated in the SAA, searched for temporal (within the T90 duration) and spatial (within the FoV) at the time of the occurrence) association with GRBs detected by satellites \cite{augu13,augu15}. 
New-Tupi is an upgrade of the previous Tupi telescopes, the detector size becomes six times larger and the plastic scintillators thickness was doubled. 
Besides the telescope mode, New-Tupi operates in the scaler mode,  where the single hit rates of all of the photomultipliers are recorded, at a frequency 0.5 Hz. 
In comparison with the telescope mode, the scaler mode of operations increases the FoV almost three fold.

In the previous Tupi cases the statistical significance of the observed signals in the counting rate associated with GRBs was above $4\sigma$. 
At least in part, this is a consequence of the very low detection energy threshold of Tupi detectors, as well as the specific physical location of the instrument in the South Atlantic Anomaly (SAA) region, where the geomagnetic field is anomalously weak $\sim 22,000$ nT.
We also noticed  that the time of an excess in the previously reported cases \cite{augu13,augu15} was near the time of the maximum of the  Earth's fair weather atmospheric electric field (the Carnegie curve) \cite{harr13}.
This time is close to the sunset time at Tupi location.
This is the time when the atmospheric conductivity is high \cite{ko10}, especially within the SAA region.
Usually these effects result in an increase in the flux of secondary charged particles, such as electrons and muons produced by primary gamma rays in the atmosphere.
Thus, more particles can reach the ground level and be registered by the detector \cite{augu15,navi17,reyes05} 

In this paper we report the results of a search for the particle excess above the background in coincidence with Fermi LAT detection of GRBs ($\Sigma E_{\gamma} > 100$ MeV) in the period from  from 8 June 2014 to 28 February 2017. 
In this period Fermi LAT detected 46 GRB events ($https://fermi.gsfc.nasa.gov/ssc/observations/types/grbs/lat\_grbs/$). 
The coordinates of 7 GRBs listed in Table~\ref{table1} were located within the FoV of New-Tupi detector at the time of the trigger occurrence.

We show in this paper that two of these seven GRBs have a possible GeV counterpart at New-Tupi detector, they are the GRB 160709A and GRB 160625B. The signal at New-Tupi in both cases are in spatial and temporal coincidence with the Fermi-LAT observations.
We present a confidence analysis for these two events, including initially a brief description of New-Tupi detector and it's alarm system to look for signals from GRBs.

\section{New-Tupi detector} 
\label{sec:detector}

The aim of New-Tupi detector is to study space weather effects driven by diverse solar transient activities\cite{augu17}.
Since 2014 New-Tupi detector is also used to search for GeV counterparts of GRBs observed by spacecraft detectors. 
New-Tupi detector is located in Niteroi city, Rio de Janeiro state, Brazil ($22^0 53'00''S,\; 43^0 06'13'W$, 3 m above sea level), within the SAA region.  

\begin{figure}
\vspace*{-4.0cm}
\hspace*{0.0cm}
\centering
\includegraphics[width=13.0cm]{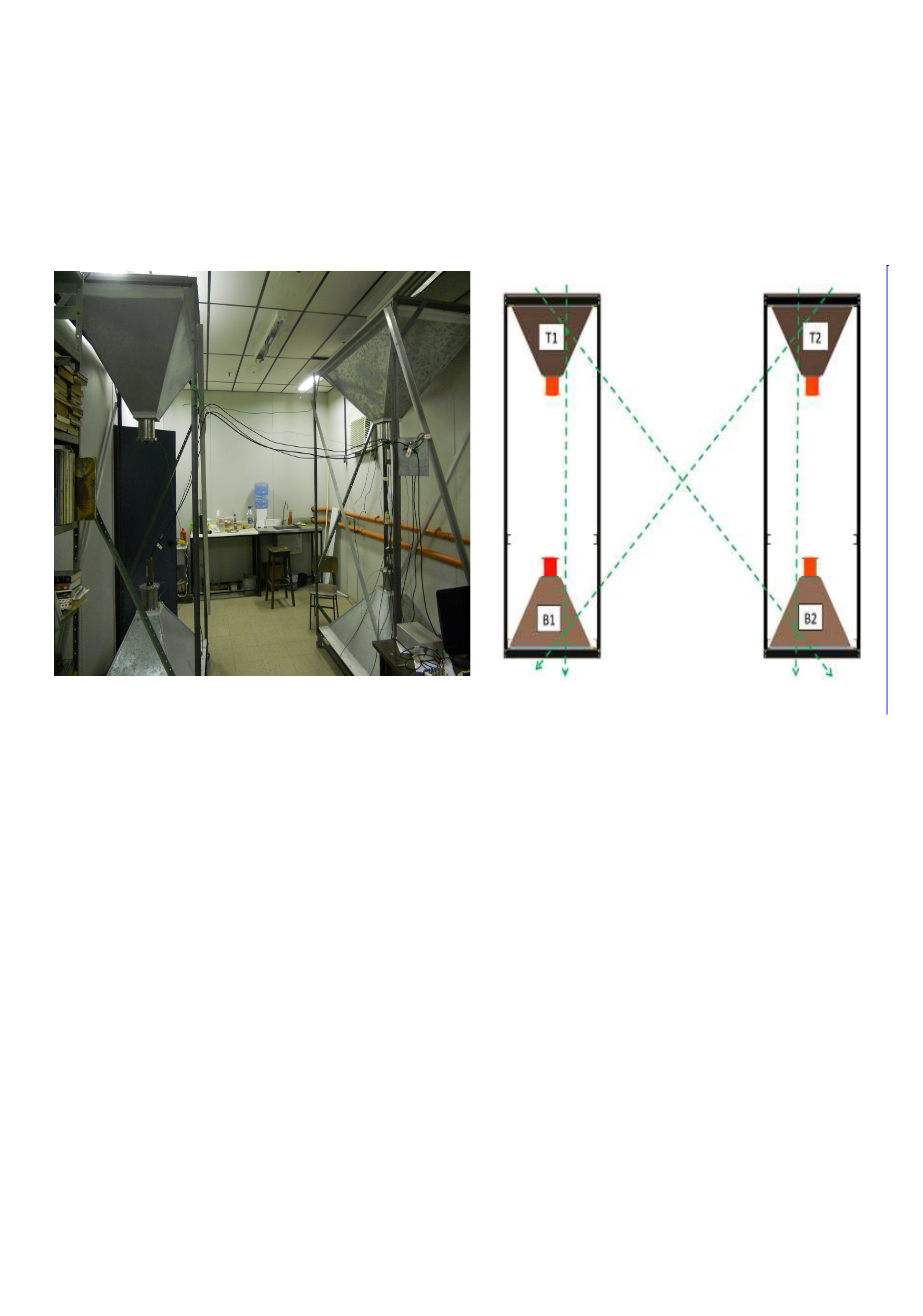}
\vspace*{-9.0cm}
\caption
{
Left: Photograph of  New-Tupi detector.
Right: Scheme of New-Tupi detector. 
In the telescope mode the counting of the coincidence pulses from two PMTs  allows to measure the secondary particle flux (air shower) from three directions, the vertical (zenith), west and east, as is indicated by the dashed lines.
In the scaler mode (no trigger) the counting rate of each PMT is recorded at a frequency 0.5 Hz.
}
\label{detector}
\end{figure}

New-Tupi detector consists of four units assembled on the basis of  of an Eljen EJ-208 plastic scintillator slab (150cm x 75cm x 5cm)  and a Hamamatsu photomultiplier (PMT) with 10 stages (model R877) of 127 millimeters in diameter.
Each unit is placed inside a box with a truncated square pyramid shape  (see Fig.~\ref{detector}).
Only signals above the threshold value, corresponding to an energy of about 100 MeV are registered. 
The output raw data of each detector is recorded at a frequency of 0.5 Hz.

\subsection{Efficiency and fiducial volume of New-Tupi detector}

 The detection efficiency of a plastic scintillator to detect charged particles such as muons
becomes fully efficient above 400 MeV \cite{alan12}.  In order to obtain the efficiency counting rate of New-Tupi detector,the 3-fold coincidence rates technique using cosmic ray muons was performed in the Muonca detector (a twin detector of New-Tupi)\cite{faut15}. The result was 
an average efficiency of 95\% for muons with energies above 100 MeV and inciding with
 a zenith's angle from 0 to 60 degrees.

In addition, the response at the periphery of the detector has poorly efficacy, 
because events that happen in this region can be losses. The events 
are detected entirely from the fiducial volume (the reliable, 
central area of the detector). The fiducial volume depend on the 
geometry of detector. In the case of a rectangular detector (scintillator)
 the fiducial volume can be obtaining  ``removing''  5 cm of the sides 
edges of the detector \cite{alle85}. For New-Tupi detector the fiducial volume is 0.9 of total volume.

\subsection{New-Tupi in the telescope mode}
\label{subsec:telescope}

Fig.~\ref{detector} (left) shows a photography of New-Tupi detector. 
The four detector units are placed in pairs, with the T1 and T2 detectors at the top, and B1 and B2 at the bottom, as shown in Fig.~\ref{detector} (right).
This layout allows to measure the secondary particle flux from three directions, the vertical (zenith), west and east.
The last two with an inclination of around $45^0$. 
The telescopes register the coincidence rate between T1 and B1; T2 and B2  (vertical), T1 and B2 (west), and T2 and B1 (east).
The horizontal and vertical distance between the corresponding units is 2.83 m.

\subsection{New-Tupi in the scaler mode}
\label{subsec:scaler}

In parallel with the telescope mode, New-Tupi is operated in the scaler mode (or single particle technique) \cite{obrian76,morello84,aglietta96}, where the counting rate of each PMT is recorded every 2 s.
This method allows to look for an increase in the particle flux counting rates, even when the source is out of the FoV of the telescopes.
This is because the FoV of a single unit is wider than the FoV of the whole detector.
However, the particle detection efficiency in the scaler mode decreases as the zenith angle of the incident particle increases, due to the atmospheric absorption. 
Thus, the scaler mode is limited to incident particles with the zenith angle less than $60^0$.

\section{The Event Rate}
\label{sec:rate}

Since June 2014  New-Tupi detector has been searching for transient events such as GeV counterparts to GRBs. 
The search includes two forms of New-Tupi data acquisition: the telescope mode and the scaler mode. 
A semi-automatic system has been developed for a cross-check between the GRB catalogs and  New-Tupi raw data logs.

\begin{figure}[!h]
\vspace*{-0.0cm}
\hspace*{1.0cm}
\centering
\includegraphics[width=14.0cm]{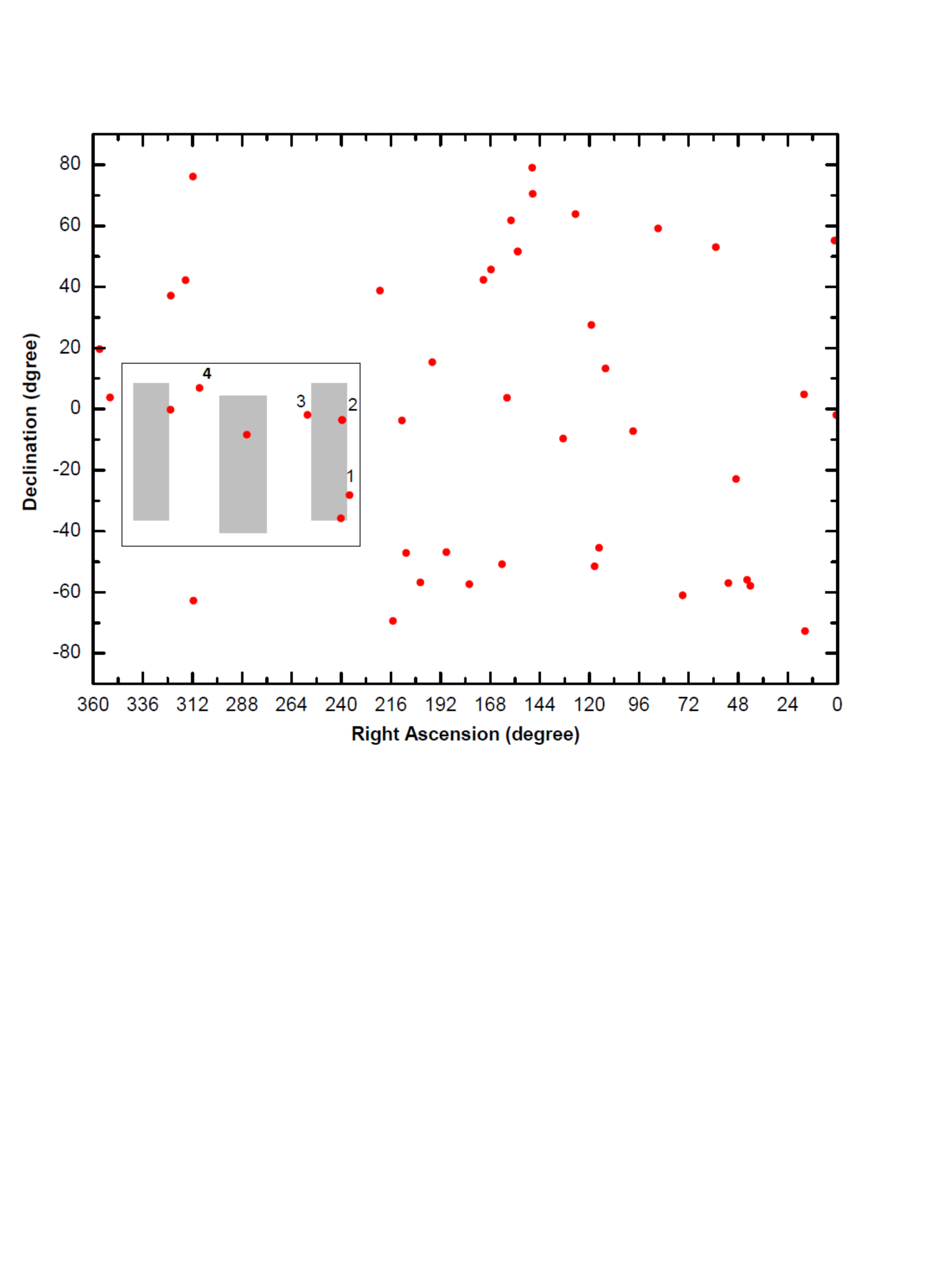}
\vspace*{-7.5cm}
\caption{
Equatorial coordinates of New-Tupi detector. The gray shaded rectangles show the FoV of New-Tupi telescopes (three units). The solid line area represents the FoV of New-Tupi in the scaler mode.  The red circles indicate the trigger coordinates of 46 GRBs ($>100$ MeV) detected by Fermi LAT in the period from 7 June 2014  to 28 February 2017.  The circles with label 1 and 4 correspond to the trigger coordinates of GRB 160709A and GRB 160625B, respectively. 
}
\label{rate}
\end{figure}

During this period (until 28 February 2017) the Fermi LAT has detected 47 GRBs with photons emission above 100 MeV.  
The trigger coordinates of seven of these Fermi Lat events were within the FoV) of New-Tupi detector (scaler mode).
Among these seven New-Tupi events, one was located within the FoV of the vertical telescope and one within the FoV of the inclined telescope pointing to the west. 
Fig.~\ref{rate} and Table~\ref{table1} summarizes basic characteristics of seven GRBs detected by the Fermi LAT and located within the FoV of  New-Tupi detector (scaler mode).
For each burst, we list the Fermi LAT detected GRB name from the GRB catalog, the trigger time, the best Fermi LAT on-ground location (Right Ascension and Declination), and New-Tupi statistical significance observed in the time profiles of the counting rates.
 The circles with label 1 and 4 in Fig.~\ref{rate} correspond to the trigger coordinates of GRB 160709A and GRB 160625B, respectively. 

\begin{table}[h]
\caption{
Basic information on GRBs located within the FoV of New-Tupi telescopes at the time of the Fermi LAT occurrence
}
\centering
 \begin{tabular}{@{}llllll} 
 \mr
GCN Name &Date  & Time  & R.A & Dec & Significance (New-Tupi)\\ 
         &    (UTC)  &     (UT)  &  (degree)   & (degree)    & (standard deviation)   \\ 
\hline
170228A & 2017-02-28	& 19:02:59.71 & 239.55 & -3.59 &  2.2   \\
170214A & 2017-02-14	& 15:34:25.92 & 256.33 & -1.88 &  2.5   \\ 
160709A & 2016-07-09	& 19:49:03.50 &	235.996& -28.188&  3.5  \\
160625B & 2016-06-25	& 22:40:15.28 & 308.597&  6.9189&  4.8 \\
150127A & 2015-01-27	& 09:32:44.14 & 285.7  & -8.4   &  2.0  \\
150118B & 2015-01-18	& 09:48:17.76 & 240.24 & -35.75 &  2.3  \\
141028A & 2014-10-28	& 10:55:03.08 & 322.602& -0.2313&  3.3  \\
 \br
\end{tabular}
\label{table1}
\end{table}

In this paper, we have obtained initially the nominal significance of the events in the time profiles, this nominal significance allow obtaining approximately  the strength of the observed signals  was calculated
according to the bin selection criteria algorithm \cite{mitr99} used in the analysis of  BATSE GRBs.
According to this
algorithm, the signal statistical significance $S$ in the ith bin is
defined as 
\begin{equation}
S_i = (C_i-B)/\sqrt{B},
\label{significance}
\end{equation}
where $C_i$ is the measured number
of counts in the ith bin and $B$ is the average background count.
This is a very popular estimation of the significance level when the 
data output is a time series and works well if the background fluctuation 
follows a Gaussian distribution.  Thus, this method maps a probability of 
a statistical fluctuation into a ``number of Gaussian sigmas''.

\section{Observations and analysis}
\label{sec:observation}

Initially, from the seven LAT GRBs whose coordinates were within the FoV of New-Tupi detector, only two of them the GRB 160709A and the GRB 160625B had a probable signal at ground level. Of them, only the signal associated with GRB 160709A was detected by the New-Tupi alert system, because it was programmed to look for excesses only before and after 100 s of the GRB's trigger. However, after a check-up, we verified that the GRB 160625B was a burst of very long duration with a peak at LAT only 181 s after the GRB trigger at Fermi GBM. Thus, we have included in this survey a straightforward analysis on signal at New-Tupi in possible association with GRB 160625B.

\section{GRB160709A}

\begin{figure}
\vspace*{0.0cm}
\hspace*{0.0cm}
\centering
\includegraphics[width=11.0cm]{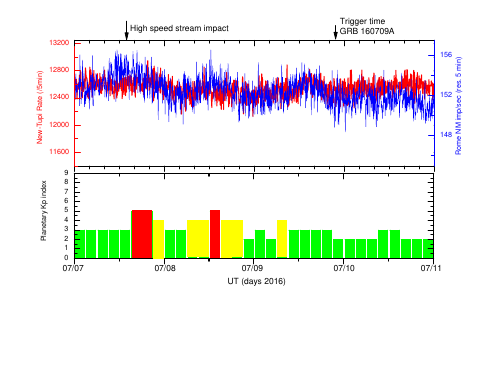}
\vspace*{-2.0cm}
\caption
{ Top panel: time profiles of the counting rate                                                                                                                                                                                                                                                                                                                                                                                                                                                                                                                                                                                                                                                                                                                                                                                                                                                                                                                                                                                                                                                                                                                                                                                                                                                                                                                                                                                                                                                                                                                                                                                                                                                                                                                                                                                                                                                                                                                                                                                                                                                                                                                                                                                                                                                                                                                                                                                                                                                                                                                                                                                                                                                                                                                                                                                                                                                                                                                                                                                                                                                                                                                                                                                                                                                                                                                                                                                                                                                                                                                                                                                                                                                                                                                                                                                                                                    in New-Tupi detector (5 min binning, scaler mode; red color) and the Rome NM (blue color) during four consecutive days (7-10 July 2016). 
Bottom panel: time profiles of the interplanetary Kp index during the same time interval.
The two vertical arrows at the top bar indicate: the impact of a high speed stream on 7 July 2016 and 
the GRB 160709A trigger time on 9 July 2016, respectively.
}
\label{rome}
\end{figure}

9 July 2016  was a typical quiet day, with no anomalous changes in the atmospheric pressure, temperature, or other known environmental parameters. Solar activity was weak, the solar  X-ray flux was quite low and the proton flux at GOES satellite remained at the background level during the day. 
The same situation was observed for the estimated 3hr planetary Kp index, that is a widely used indicator (an integer in the 0-9 range; indices of 5 or greater indicate storm-level) of the mean magnetospheric activity.
However, around two days before, there was registered a high speed stream (HSS) impact on the Earth magnetosphere, as shown in Fig.~\ref{rome}. 
The impact produced a $Kp=5$ (minor) geomagnetic storm. 
The duration of the magnetic perturbation was more than 30 hours, as can be seen in Fig.~\ref{rome} (bottom panel). 
The effects of this minor geomagnetic storm on the ground level particle detectors was a small fall in the counting rate, the so-called Forbush decrease. 
We present in Fig.~\ref{rome} (top panel) a comparison between the counting rate between two ground level detectors,  New-Tupi vertical telescope, and the Rome NM \footnote{The ground based Neutron Monitors (NM) around the world are designed to measure the intensity and variation of galactic cosmic rays.
The Earth based observations of space weather effects are complementary to the satellite measurements.} . 
Both detectors are positioned at sea level, located at equivalent latitudes (in southern and northern hemispheres, respectively), and have similar geomagnetic rigidity cutoff thresholds. 

According to Guiriec et al. (GCN 19675) on  9 July 2016 at 19:49:03.50 UT the Fermi LAT detected GRB 160709A, which was also detected and triggered by Fermi GBM (trigger $489786546 / 160709826$).
The best LAT on-ground location was found to be (R.A., Dec.) = (236.110, -28.500) deg. 
The Fermi LAT data show a significant increase in the event rate in the 30-100 MeV energy range in spatial and temporal correlation with the Fermi GBM prompt emission (in the keV energy range). In addition, more than 30 photons above 100 MeV are observed within 100 seconds, the highest energy photons being 1 GeV events observed 2 seconds
after the GBM trigger and consistent with the GBM brightest emission episode.

Also on 9 July 2016  New-Tupi alert system working in the scaler mode detected an excess in the particle counting rate in association with the prompt emission of GRB 160709A.
From the GRB 160709A trigger coordinates at Fermi LAT we can deduce that the arrival direction of GRB made an angle of $\sim 45^0$ relative to the vertical direction(zenith) at New-Tupi location.  New-Tupi onset time was delayed by 1.2 seconds with respect to the GRB 160709A trigger time.

\begin{figure}
\vspace*{0.0cm}
\hspace*{0.0cm}
\centering
\includegraphics[width=12.0cm]{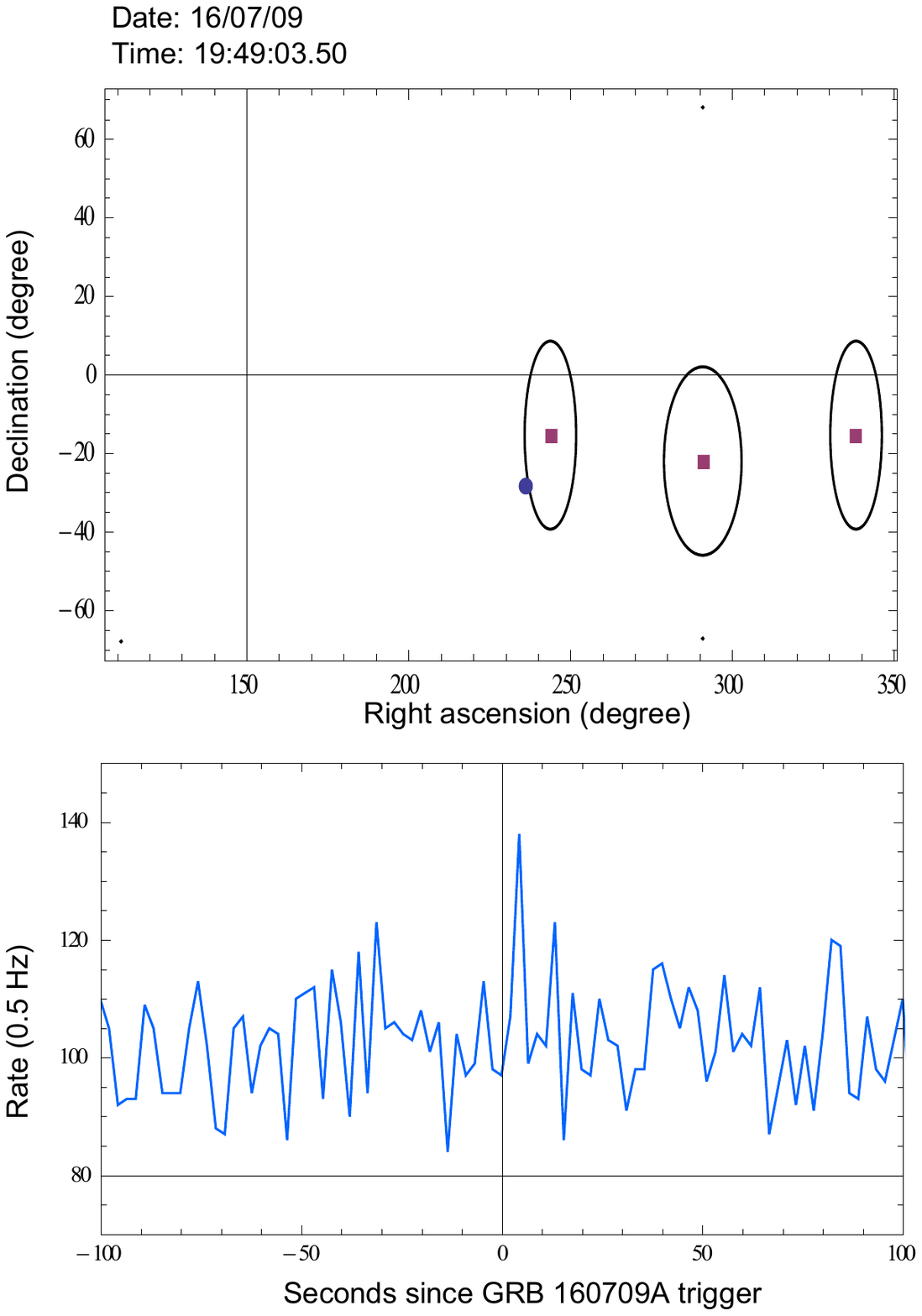}
\vspace*{-5.0cm}
\caption
{
The output of the New-Tupi alert system on 9 July 2016. 
Top panel represents the position of  New-Tupi telescopes in equatorial coordinates, ellipses with square dots demonstrate approximately the effective field of view of each telescope, and the solid circle indicates the detected position (coordinates) of the Fermi LAT. 
Bottom panel shows New-Tupi counting rate                                                                                                                                                                                                                                                                                                                                                                                                                                                                                                                                                                                                                                                                                                                                                                                                                                                                                                                                                                                                                                                                                                                                                                                                                                                                                                                                                                                                                                                                                                                                                                                                                                                                                                                                                                                                                                                                                                                                                                                                                                                                                                                                                                                                                                                                                                                                                                                                                                                                                                                                                                                                                                                                                                                                                                                                                                                                                                                                                                                                                                                                                                                                                                                                                                                                                                                                                                                                                                                                                                                                                                                                                                                                                                                                                                                                                              recorded at a frequency 0.5 Hz in the scaler mode, 100 s before and 100 s after the LAT onset time.
}
\label{task}
\end{figure}

The nominal statistical significance of New-Tupi counting rate excess was estimated as $3.45 \sigma$.
We summarize the situation in Fig.~\ref{task}, where the output of the alert system is presented.

\begin{figure}
\vspace*{0.0cm}
\hspace*{0.0cm}
\centering
\includegraphics[width=10.0cm]{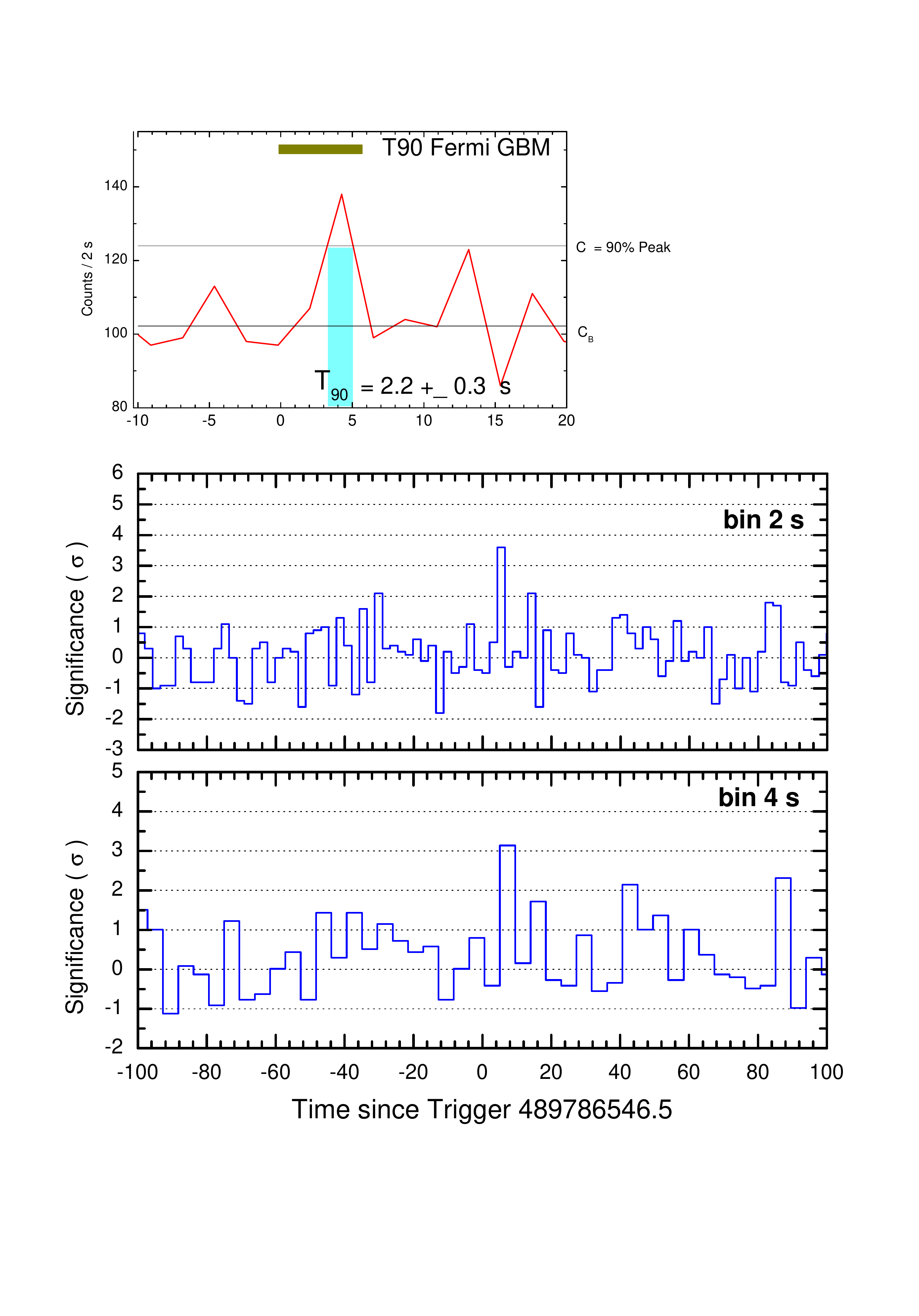}
\vspace*{-1.5cm}
\caption
{
Top panel: the counting rate observed in New-Tupi telescope in the scaler mode on 9 July 2016.
The shaded area illustrates the T90 duration of the signal associated with the prompt emission of GRB 160709A
at Fermi GBM. 
Central and bottom panels: statistical significance (number of standard deviations) of the 2 s and 4 s binning counting rate observed by New-Tupi
telescope inside a 200 s interval (100 s before and 100 s after the Fermi LAT onset time).
}
\label{refine}
\end{figure}

A more detailed picture is shown in Fig.~\ref{refine} (top panel), where the duration parameter $T_{90}$, defined as the time interval over which 90\% of the total background-subtracted counts are observed, is estimated as $2.2 \pm 0.2$ s.

Statistical significance as above defined in the 2 s and 4 s binning counting rates inside a 200 s interval (100 s before and 100 s after the trigger) observed by New-Tupi detector as a function of the time elapsed since the GRB 160709A trigger time can be seen in Fig.~\ref{refine} (central and bottom panels).
According to Eq.~\ref{significance}, a signal (particle excess) with a significance of up to $3.66\sigma$ was found in the raw data.

\begin{figure}
\vspace*{-2.0cm}
\hspace*{0.0cm}
\centering
\includegraphics[width=13.0cm]{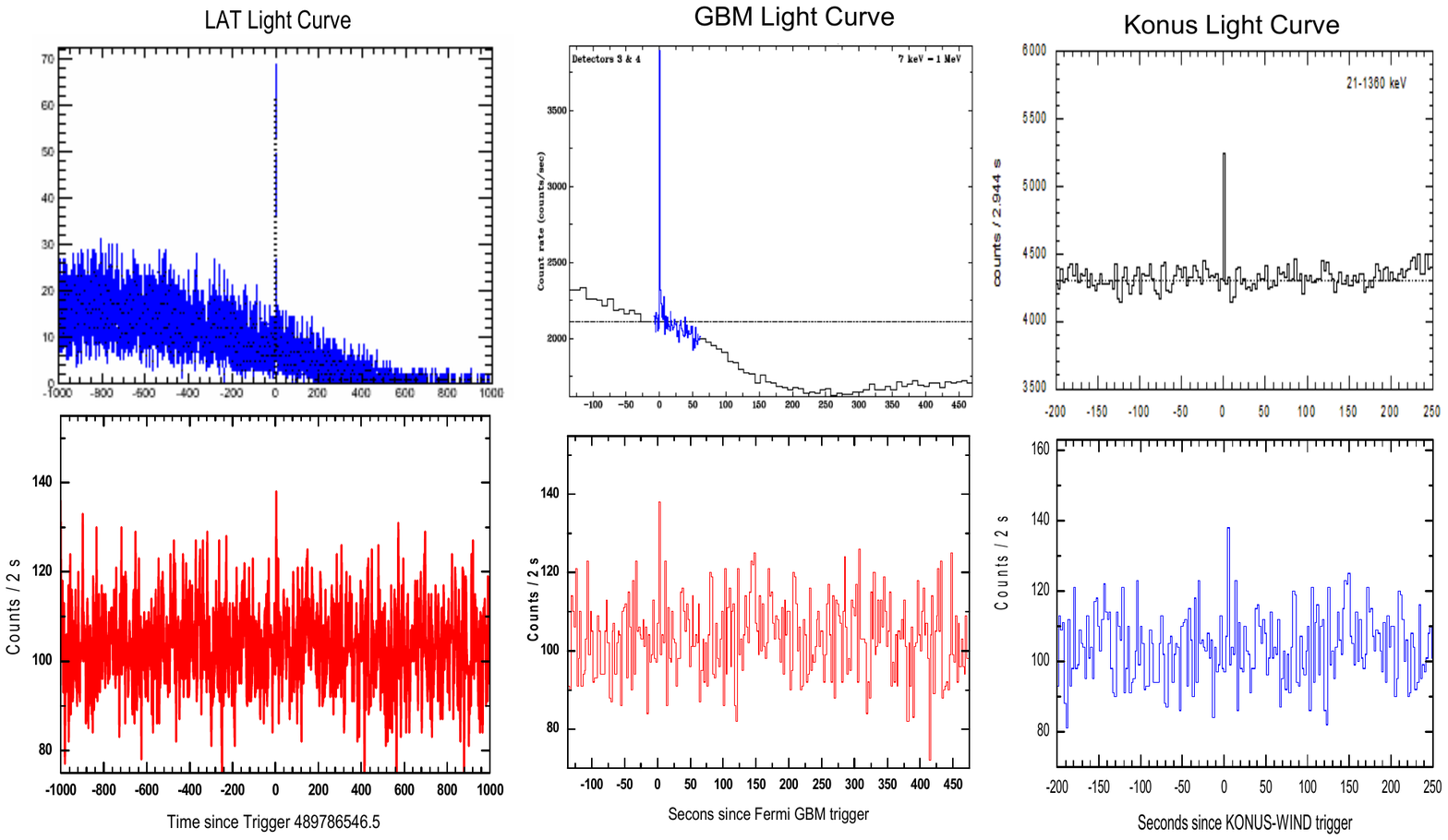}
\vspace*{-8.0cm}
\caption
{
Top:(Left panel) Fermi LAT low energy light curve, between $30-100$ MeV energy band, associated with the prompt emission phase of GRB 160709A. (Central panel) Fermi GBM light curve, between $50_300$ keV energy range associated with GRB 160709A. (Right panel) Konus-Wind light curve, from 20 KeV to 10 MeV energy band, associated with GRB 160709A
Bottom panels: In all cases, the time profile of the particle counting rate (scaler mode) in New-Tupi detector.
}
\label{lat_gbm_k}
\end{figure}

A comparison between the Fermi LAT low energy light curve, between $30-100$ MeV energy range, associated with the prompt phase of GRB 160709A and the counting rate observed by New-Tupi detector is presented in Fig.~\ref{lat_gbm_k}
(left panel).
In addition, Fermi LAT has detected around of 30 events with energies above 100 MeV, and four events have energies around 1 GeV and were detected at 1,47, 1.76, 3.83 and 14.44 seconds after the GBM trigger time.

The prompt emission of GRB 160709A was also detected by Fermi GBM (trigger 489786547 / 160709826) as reported by Jenke (GCN 19676). 
The GBM on-ground location was consistent with the LAT position.
The GBM light curve consists of two peaks with a duration (T90) of about 5.6 s (50-300 keV) beginning 64 ms before the trigger time.
The Fermi GBM event fluence was estimated as $(3.33 \pm 0.09)\times 10^{-6} erg/cm^2$ in the 10-1000 keV band. 
Fig.~\ref{lat_gbm_k} (central panel) shows a comparison between the time profiles of the Fermi GBM light curve and the counting rate of New-Tupi detector associated with GRB 160709A.


According to Frederiks et al. (GCN 19677),  Konus-Wind located at Lagrange Point L1 also triggered GRB 160709A.  
The Konus light curve shows a short, hard pulse with a total duration of $\sim 0.4$ s. 
The emission was seen up to $\sim 8$ MeV and a fluence of $(1.5 \pm 0.3) \times 10^{-5} erg/cm^2$ in the $20 keV - 10 MeV$ energy range. 
Fig.~\ref{lat_gbm_k} (right panel) shows a comparison between the time profiles of Konus-Wind and New-Tupi.


The bright short-duration  GRB 160709A was detected by several other space based detectors. 
At  19:49:03.464 UT on 9 July 9 2016, Swift-BAT detected a count rate increase and a sub-threshold signal in the image reported by Sakamoto et al.(GCN 19681). 
The CALET gamma-ray burst monitor on board of the ISS triggered GRB 160709A at 19:49:04.67 (Asaoka et al. GCN 19701).
A Cadmium-Zinc-Telluride coded-mask imager (CZTI) of  the multi-wavelength astronomy mission ASTROSAT on a 650-km near-equatorial orbit, detected a single peak at 19:49:04.00 UT,  0.5 seconds after Fermi GBM Trigger (Bhalerao et al. GCN 19740).

There are several reports on the GRB 160709A afterglow observations. 
The MASTER II  robotic telescope located in South African Astronomical Observatory (SAAO) was pointed to  GRB160709A at 19:50:49 UT, 66 sec after notice time and 105 sec after trigger time (Lipunov et al. GCN 19678).
Swift-XRT has performed a series of observations, tiled on the sky. 
The total exposure time was 3.3 ks. 
However, the XRT data have been overlapped with previously catalogued X-ray sources, so their status made them unlikely to be the afterglow. (Roegiers et al. GCN 19679).





\subsection{Effective significance}

We have begun the analysis examining the counting rate profile, as shown in 
Fig.~\ref{refine} using the nominal significance level 
(in units of standard deviations) as defined in Eq.~\ref{significance} for the two hours interval, one hour before and one hour after the GRB160709A trigger time.
In this time interval, we have found only 
an event with a nominal significance of $3.5\sigma$  and this event is linked with the GRB 160709A.

However, in order to make a more precise analysis on the confidence with which the signal associated with GRB160709A is observed at New-Tupi detector. We make an analysis on the background fluctuation on the counting rate to a period of 24 hours. 

\begin{figure}
\vspace*{-0.0cm}
\hspace*{0.0cm}
\centering
\includegraphics[width=10.0cm]{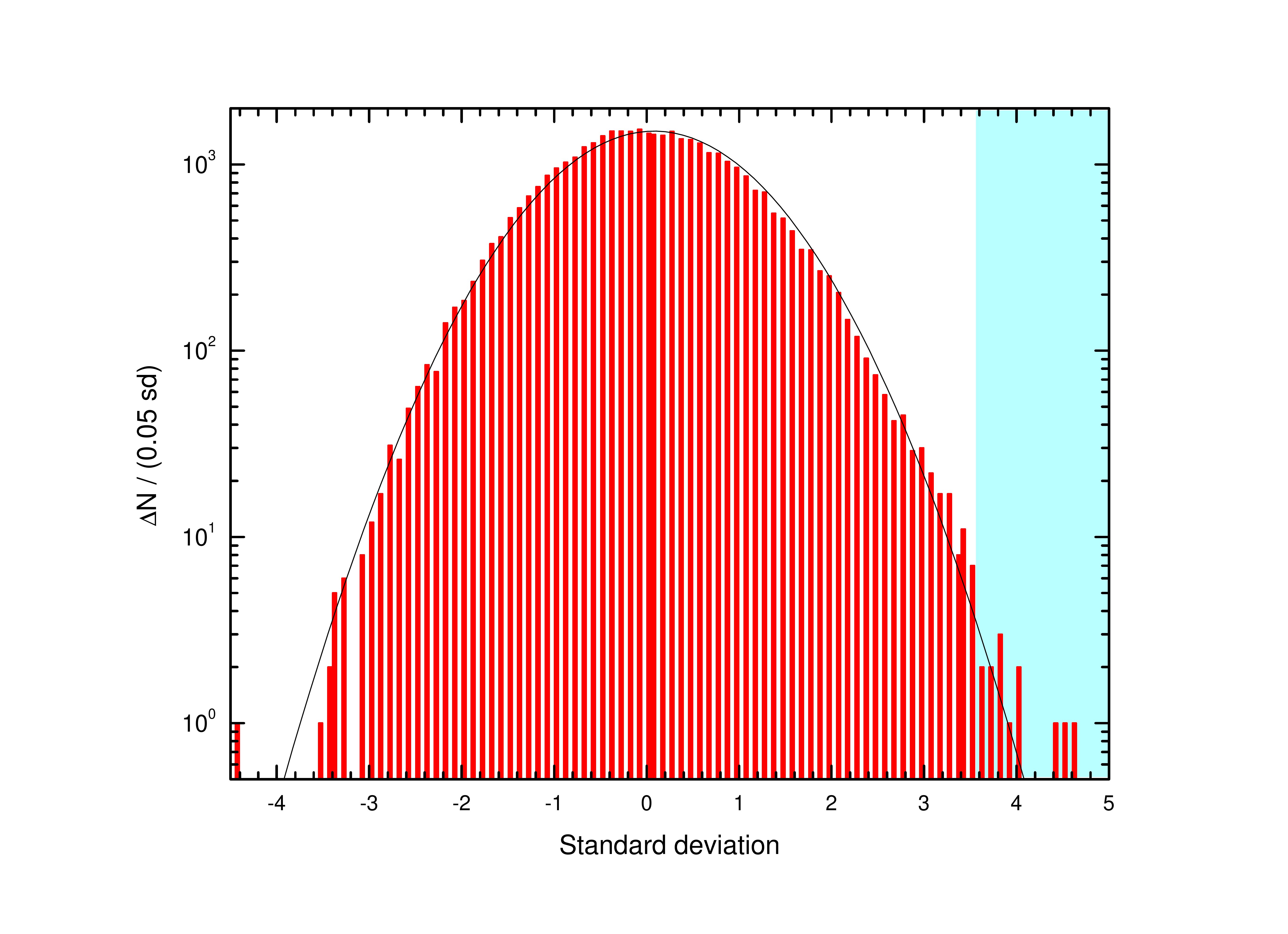}
\vspace*{-0cm}
\caption
{
Distribution of the fluctuation counting rate (in units of standard deviations), in bins of $0.05\sigma$ for New-Tupi detector in the scaler mode,  inside the 24 hours interval (on 9 July 2016). 
 The light blue band at right, mark the region of  13 events with a significance equal or above $3.5\sigma$ observed on 9 July 2016.
The solid curve represent a  Gaussian fit distribution.
}
\label{gauss2}
\end{figure}

Assuming there are no astrophysical signals in the New-Tupi data, we can
search through 24 hours of data, i.e, in each of the 43,200 bins registered in one day, 
is determining the counting rate and its significance (using Eq-\ref{significance}) and enter into a histogram. Fig.~\ref{gauss2} summarizes the situation for 9 July 2016 day.

The signal in time profiles of New-Tupi associated with the GR 160709A has a nominal significance of $3.5\sigma$,
this signal is within the T90(=5.6 s) GRB duration, as observed by Fermi GRM (see top panel of Fig.~\ref{refine}).
However, following Fig.~\ref{gauss2}, we can see that there are 13 events with significance equal or above
$3.5\sigma$, they are within the light blue band.  We would like to 
point out, that the number the events with a significance above $3.5\sigma$ is in average 4, in a quiet day (without solar storms). The high number of events with significance above $3.5\sigma$ on July 9, 2016, is due to the fact that this day was not a calm day. Indeed, 24 hours earlier there was a small geomagnetic storm, as shown in Fig.~\ref{rome}.

Thus, the probability of we have 13 events with a nominal significance equal or above
$3.5\sigma$ per bin in one day (on 9 July 2016), is $P_{13}(=13/43200=3.00\times 10^{-4})$.
Then, the probability of one or more events with a nominal significance equal or above $3.5\sigma$, to be within 
the duration T90 (= 5.6 s), and this is equivalent to $\sim  n(=3)$ bins, can be obtained
using the binomial probability distribution, whose mean value is defined as
\begin{equation}
\mu=nP_{13}=9.00\times 10^{-4}.
\end{equation}
This probability, is equivalent to a Gaussian significance
of $3.0\sigma$.


However, there is also some favourable aspects, such as a good temporal coincidence, the New-Tupi event is within the T90 duration ($\sim 5$ s) of GRB 160709A observed by Fermi GBM and the probability to happen this is $p= 5/86400=5.79 \times 10^{-5}$. In addition, we have verified that in the time interval of two hours 
(one hour before and one hour after of Fermi GBM trigger time) there is only one event with a nominal significance of $\sim 3.5\sigma$, that is, the New-Tupi event associated with GRB160709A.

\subsection{Probability of false alarm detection}

Considering that, during the period from June 2014 to February 2017, only 7 LAT GRBs were within the field of view of the New-Tupi detector, that is, we analyzed 7 different LAT GRBs (N = 7), and found only one event with a nominal confidence of 
$3.5\sigma$ with a chance to be associated with a LAT GRB. This information allows us to obtain the probability $P_{FD}$ that this event be a false detection as 
\begin{equation}
P_{FD}=(1-p)^N=0.9937,
\end{equation}
where $ p\sim 9.00 \times 10^{-4}$ is the average probability of one or more events with a nominal significance equal or above $3.4\sigma$, to be within the T90($=5.6$ s) duration
of the GRB 160709A as observed by Fermi GBM.
Consequently, the probability of not be a false detection is

\begin{equation}
1-P_{FD}=1-(1-p)^N=6.28 \times 10^{-3}.
\end{equation}

\section{GRB160625B}

June 25,  2016, was the third day in a row, with the face of the sun is blank, no sunspots. As a result, solar activity was low and according to NOAA the chance of a strong solar activity was no more than 1\%. These were quiet days, without solar storms.

According to  Dirirsa et al. (GCN 195860) on 25 June 2016 at 22:40:15.28 UT the Fermi LAT
detected GRB 160625B which was also detected and triggered by Fermi GBM (trigger 
$488587880 / 160625952$). The best LAT on-ground location is found to be:
(R.A., Dec.) = (308.56, 6.93) deg. 

LAT has detected more than 300 photons with energies above 100 MeV.  The LAT emission consists of the main peak at $\sim T0+181$ s. However, the GRB was detectable by LAT up to $\sim 1$ ks after the trigger.  The highest-energy photon was a 15 GeV event which was observed at $\sim 345$ s after the GBM trigger.

Initially, the New-Tupi alert system did not detect this event, because it was programmed to look for excesses
only 100 s before and 100 s after trigger time. The circle with label 4 in Fig~\ref{rate}
show the GRB 160625B trigger coordinates, it is within of field of view of  New-Tupi detector. The incident angle of the GRB 160625B relative to the zenith of New-Tupi location was only of 29 degrees. 

\begin{figure}
\vspace*{0.0cm}
\hspace*{0.0cm}
\centering
\includegraphics[width=17.0cm]{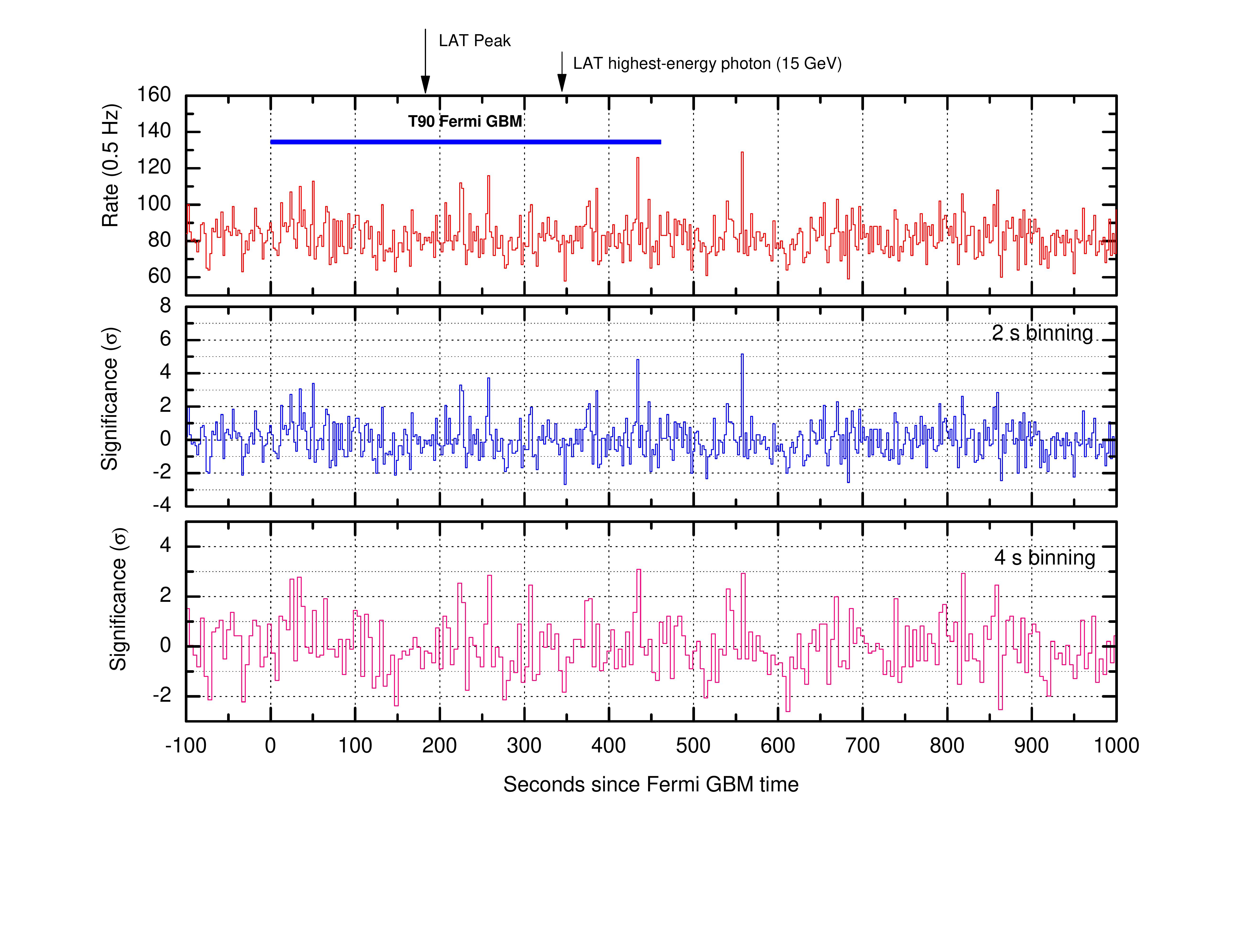}
\vspace*{-2.5cm}
\caption
{
Top panel: the counting rate observed in New-Tupi telescope in the scaler mode on 25 June 2016. The horizontal narrow blue band
illustrates the T90 duration associated with the GRB 160625B
at Fermi GBM. 
Central and bottom panels: nominal statistical significance (number of standard deviations) of the 2 s and 4 s binning counting rate observed by New-Tupi
telescope inside a 1100 s interval (100 s before and 1000 s after the Fermi GBM trigger).
}
\label{160625B}
\end{figure}

Fig.~\ref{160625B} show the time profiles of New-Tupi counting rate (top panel) and
the statistical significance, as above defined, in the 2 s and 4 s binning counting rates inside an 1100 s interval (100 s before and 1000 s after the trigger) observed by New-Tupi detector as a function of the time elapsed since the GRB 160625B trigger time can be seen in Fig.~\ref{160625B} (central and bottom panels)

The two peaks observed in New-Tupi time profiles, with a significance of about 
$4.8\sigma$ and $5\sigma$ are located at 438 s, and 558 s after the GBM trigger, as shown Fig.~\ref{160625B}. Of these two peaks, that with a significance of $4.8\sigma$ is within the T90 duration of Fermi GBM. These two peaks persist even at 4 s binning counting rates, with a significance of about $3\sigma$. In addition, there about 6 events with a significance equal or above  $3\sigma$ (2 s binning) within the T90 duration.

\begin{figure}
\vspace*{0.0cm}
\hspace*{0.0cm}
\centering
\includegraphics[width=17.0cm]{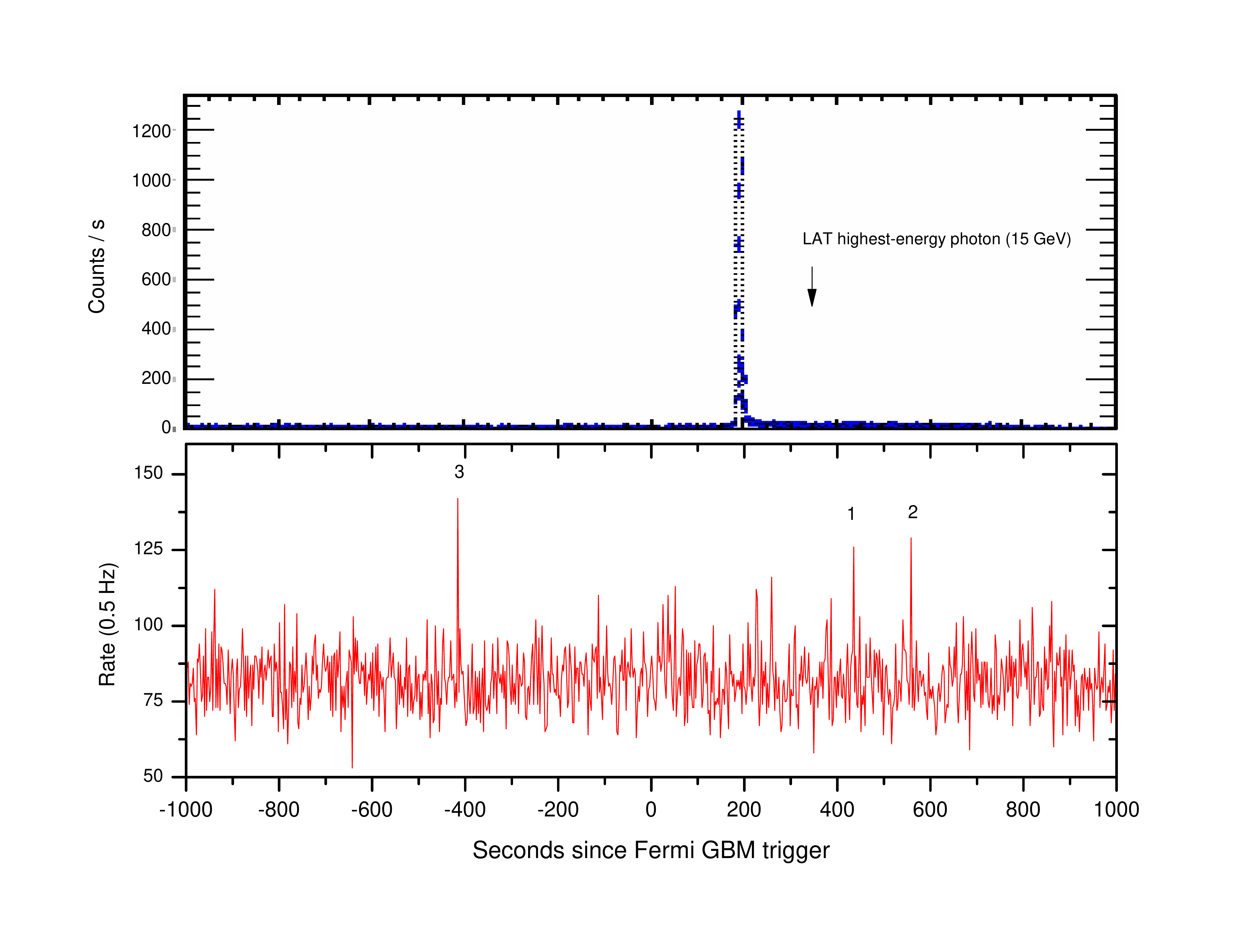}
\vspace*{-1.5cm}
\caption
{
Top panel: LAT light curve, between $30-100$ MeV energy band, associated with the GRB 160625B. Bottom panel: time profiles of the particle counting rate (scaler mode) in New-Tupi detector. Inside 2000 s interval (100 s before and 1000 s after the Fermi GBM trigger.
}
\label{LAT_curve}
\end{figure}

Fig.~\ref{LAT_curve} shows a comparison between the light curve observed by Fermi LAT and the time profiles (counting rate) observed by New-Tupi, inside a 2000 s interval (1000 s before and 1000 s after the trigger). 
Besides the two peaks with a significance above $4.8\sigma$ already mentioned above
(labeled as 1 and 2), there is at New-Tupi another peak, 415 s before
of the trigger time, with a nominal statistical significance above $6\sigma$ (labeled as 3). At this moment we do not have conditions to confirm if this event could be a precursor signal from GRB 160625B.  

\subsection{Effective significance}

\begin{figure}
\vspace*{0.0cm}
\hspace*{0.0cm}
\centering
\includegraphics[width=17.0cm]{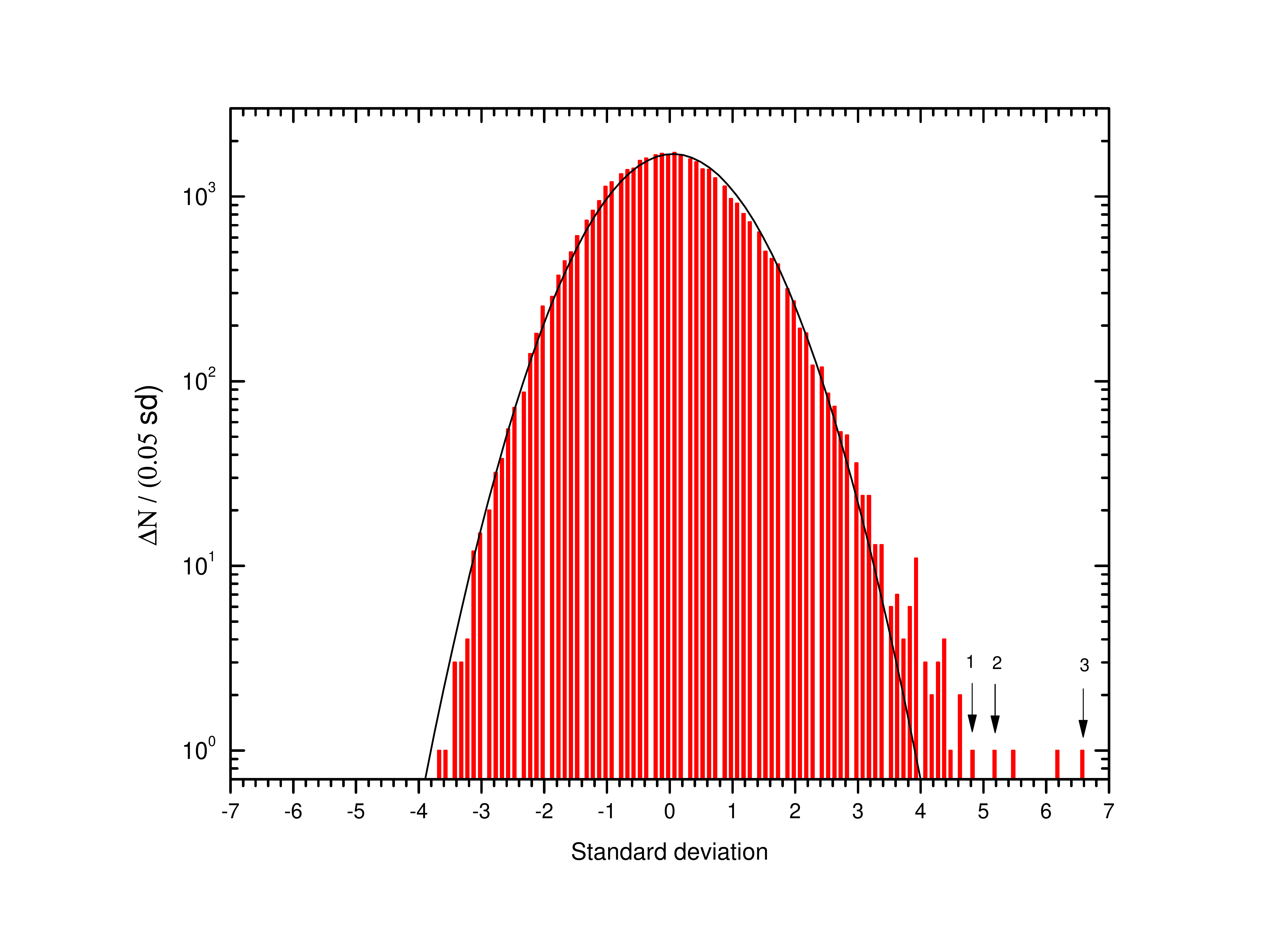}
\vspace*{-1.5cm}
\caption
{
Distribution of the fluctuation counting rate (in units of standard deviations), in bins of $0.05\sigma$ for New-Tupi detector in the scaler mode,  inside the 24 hours interval (on 25 June 2016). 
The vertical arrows labeled as 1, 2 and 3 indicate those events that appear in 
Fig.~\ref{LAT_curve}.
The solid curve represents a  Gaussian fit distribution.
}
\label{gaussB}
\end{figure}

We have examined the time profiles observed by New-Tupi detector (scaler mode) in a time interval of 24 h on 25 June 2016, using the statistical significance level distribution (in units of standard deviations), using the 2 s binning counting rate, and it was determined using Eq.~\ref{significance}.
The result is the histogram, plotted in $0.05\sigma$ bins, a Gaussian distribution as shown in Fig.~\ref{gaussB}. The events labeled as 1, 2 and 3 are those
with a nominal significance above $ 4.8 \sigma$ and are the same ones that appear in 
Fig.~\ref{LAT_curve}.

\begin{figure}
\vspace*{0.0cm}
\hspace*{0.0cm}
\centering
\includegraphics[width=17.0cm]{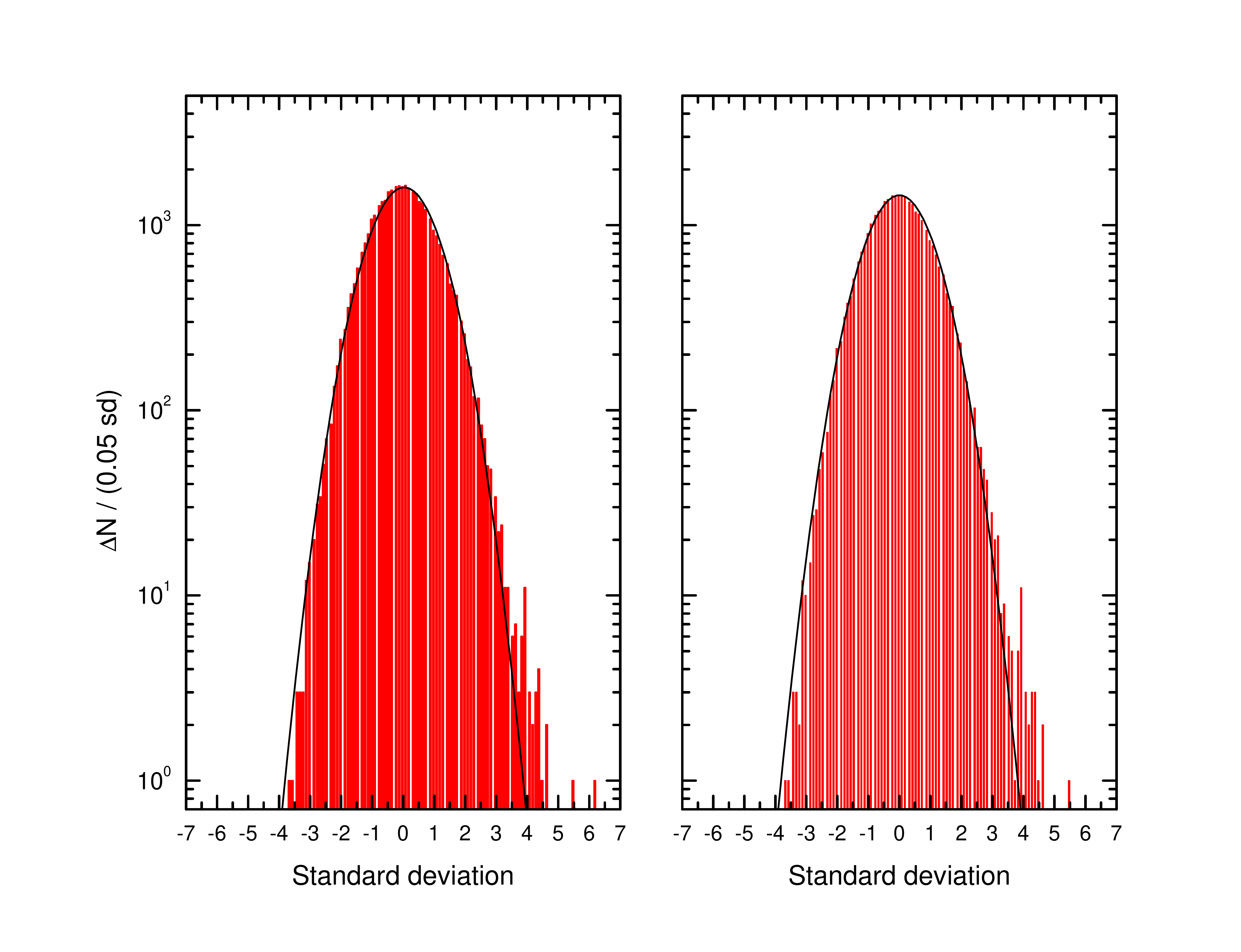}
\vspace*{-1.5cm}
\caption
{
The same as in Fig.~\ref{gaussB}. However, after excluding the counting rate $\pm 2000$ s (left panel) and $\pm 2500$ s (right panel) around the trigger time.
}
\label{gaussC}
\end{figure}

The vertical arrows labelled as 1, 2 and 3 in Fig. 10, represent the events marked also in Fig. 9. Of those, the event labelled as 1 is within the T90 duration of GRB, as observed by Fermi GBM. We can see that they
are out of the Gaussian curve, that covers the background fluctuation. However,
a confidence analysis to obtain the effective statistical significance require a careful analysis because of the GRB 160625B has a very long duration, the high energy photons were emitted during an elapsed time of more of 1 ks.
In this paper, only a straightforward analysis is made in order to obtain the effective significance.

With this aim, we have excluded from the analysis the counting rate observed 2000 s before and after the GRB trigger and the new distribution of the counting rate is shown in Fig.~\ref{gaussC} (left panel). From this figure, we can see that still there are two events with a nominal significance above $4.8\sigma$. They are exactly the two events which were not labeled in the Fig.~\ref{LAT_curve}. Increasing the exclusion to $\pm 2500$ s around the nominal trigger, the number of events with significance above $4.8\sigma$ not labeled fall to only one, as shown in Fig.~\ref{gaussC} (right panel).

Taking into account a background rate of 2 events with a significance equal to or above $4.8\sigma$ per bin in one day, this probability is $P_2(= 2/43200 = 4.63 \times 10^{-5})$. Then, the
probability of one or more events with a nominal significance equal or above $4.8\sigma$, within the duration T90 (= 460 s), or n(= 230) bins, can be obtained using the binomial probability distribution, whose mean value is defined as
\begin{equation}
\mu=nP_2=1.0 \times 10^{-2}.
\end{equation}
This probability is equivalent to a Gaussian siginificance of $\sim 2\sigma$.

In addition,
the probability of observing in a period of 24 h (on 25 June 2016), an excess in the New-Tupi counting rate, within the range $ T90 = 460$ s  duration of the GRB 160625B as observed by Fermi GBM, 
be a random coincidence is estimated as $p=460/86400=5.32\times 10^{-3}$. 

On the other hand, as in the previous case, we consider that, during the period from June 2014 to February 2017, only 7 LAT GRBs were within the field of view of the New-Tupi detector, that is, we analyzed 7 different LAT GRBs (N = 7), and considering also only one event with chance to be associated with a LAT GRB,
the probability of the event be a false detection is
\begin{equation}
P_{FD}=(1-p)^N= 0.932,
\end{equation}
where $p=1.0\times 10^{-2}$ is the average probability of obtaining one or more events with a nominal significance equal to or above $4.8\sigma$. Then, the probability of not to be a false detection is
\begin{equation}
1-P_{FD}=6.79 \times 10^{-2}.
\end{equation}

\section{Signal-to-noise ratio}
\label{sec:snr}

The detection of gamma rays from space at ground level is done indirectly through the detection of secondary particles (air-shower)  produced in the gamma ray interactions in the atmosphere. 
This detection depends on many factors, such as the energy spectrum of the gamma rays, the duration of the burst, the altitude and the magnetic coordinates, the detector characteristics, such as the size of the detector, the detector efficiency and 
the energy detection threshold, as well as the atmospheric conditions at the time of observation.

In order to evaluate the signal strength relative to the background for the case of GRB160709A counterpart observed at ground level, we'll try to get information about the effects of the location of the detector, such as its altitude and its local magnetic coordinates.

We have analyzed via Monte Carlo method the above effects at two experimental sites,  the HAWC at Sierra Negra Mountain (Mexico) and  New-Tupi within the SAA region (Brazil).
These sites have different altitudes of observation and different geomagnetic rigidity cutoff values.
Table~\ref{table2} summarizes the geomagnetic parameters at these two locations.

The magnetic effect affects the development of an air-shower, even in the case of neutral primary particles, such as gamma-rays. 
This is so-called ``magnetic lateral dispersion'' of charged particles in an air shower.
It occurs during propagation of particles in the atmosphere. 
The component of the geomagnetic field perpendicular to the particle trajectory is the primary cause of this effect. 
This effect results in decrease of the number of particles collected by the detector, and therefore the detector sensitivity. 
The lower the transverse geomagnetic field at the experiment location, the higher the detector sensitivity. 
However, this effect also depends on the inclination of the air shower axis.

The CORSIKA Monte Carlo package \cite{heck} includes as a default the magnetic lateral dispersion of the charged particles in a shower.
For detailed calculations of particle transport and interactions with matter, in the present analysis we used the simulation code FLUKA (FLUktuierende KAskade).
\cite{ferrari05}
\begin{table}[h] 
\caption{Geomagnetic parameters at two sites used in this study}
\centering
 \begin{tabular}{cccccc} 
 \hline\hline
 Site & (Lat, Lon) & B(horizontal)   & B(vertical) & Altitude  & Rigidity \\ 
      & (degree)   &     (mT)        &      (mT)            & (m.a.s.l) & cutoff (GV)        \\
 \hline\hline
 HAWC      &(19.0N,97.3W) & 27.41 &  29.36 & 4100 & 8.0   \\
 New-Tupi    &(22.9S,43.1W) & 16.52 & -16.60 & 5 & 9.2 \\
 \hline
\end{tabular}
\label{table2}
\end{table}

\begin{figure}
\vspace*{0.0cm}
\hspace*{0.0cm}
\centering
\includegraphics[width=10.0cm]{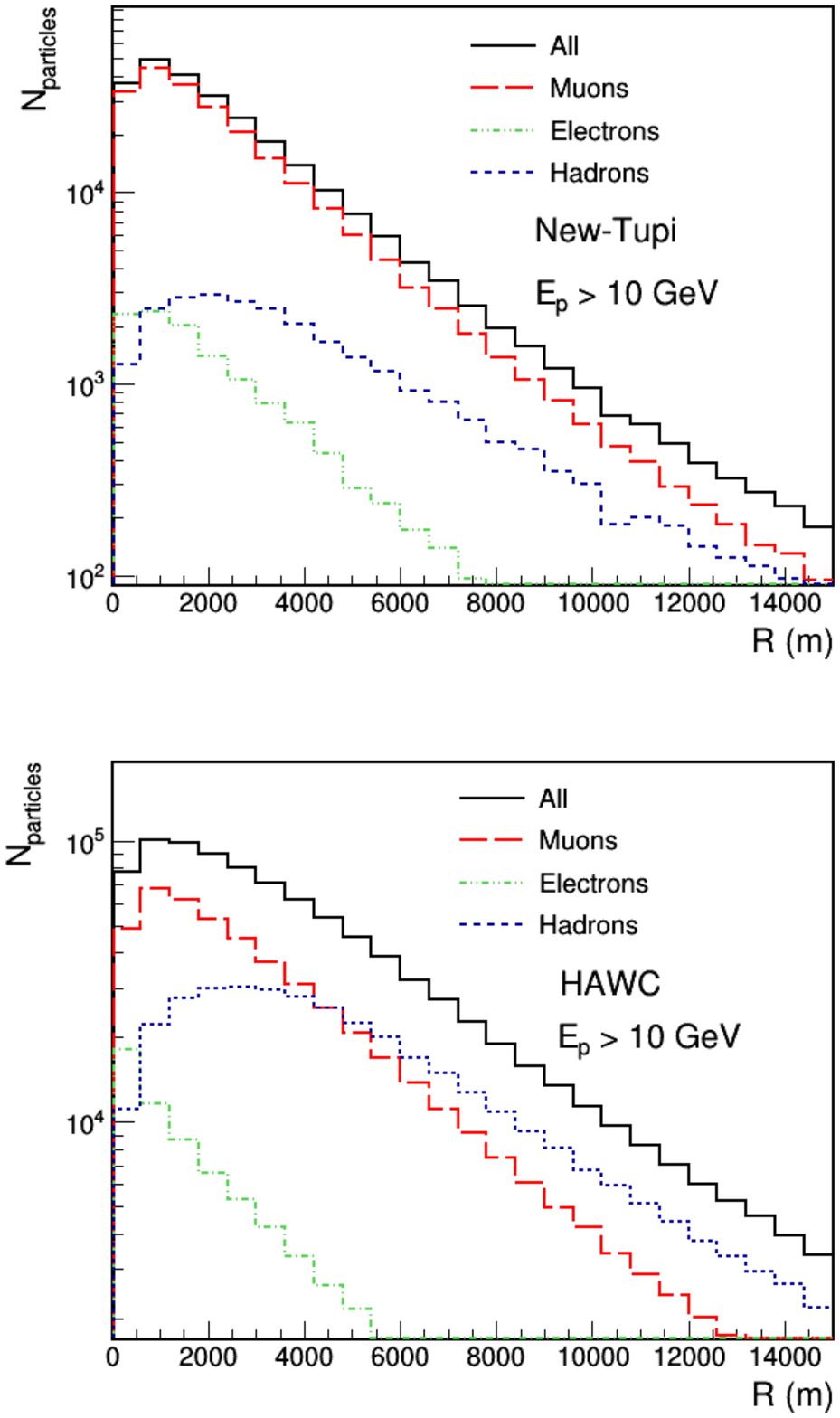}
\vspace*{-1.0cm}
\caption
{
The lateral distribution of secondary particles (muons, electrons, and hadrons) with energies above 100 MeV at ground level obtained from
a Monte Carlo simulation using $1.2 \times 10^6$ primary protons with energies above 10 GeV and a power law differential spectrum ($E_p^{-2.8}$) incident upon the atmosphere isotropically with zenith angle from $0^0$ to $60^0$. 
Top panel: New-Tupi location (3 m above sea level). 
Bottom panel: the HAWC experiment location (4100 m above sea level).
Solid lines: all secondaries. 
Dash lines: muons. 
Dot-dash lines: electrons. 
Dot lines: hadrons.
}
\label{proton}
\end{figure}

\subsection{Particle background}

The detection of a particle shower at ground level depends on a number of factors, including the trigger conditions and the effective area of the detector.

In the scaler mode (or single particle technique) \cite{obrian76,morello84,aglietta96}, the single hit rate of each detector (PMT) is recorded at a given frequency.
This mode allows a search for an excess in the counting rate.
In the scaler mode a field of view is larger and the primary particle energy is lower than those detected in the shower mode. 
However, there is no exact information on the arrival direction. 

The effective FoV in the scaler mode covers a solid angle of $\sim \pi$ sr, around the vertical direction.   
The effective angular aperture of the detector operating in the scaler mode has a zenith angle up to $60^0$.
Due the effects of absorption in the atmosphere, at the large zenith angles ($\theta > 60^0$) the Earth’s curvature must be considered.

Typically it is expected that detectors at mountain level are better suited for the detection of secondary particles from primary gamma rays. 
However, there is a side effect.
This is a high noise from cosmic ray air shower background events.
Usually the signal-to-noise ratio (S/N) is defined as a measure of signal strength relative to background noise. 
Since  $S/N \propto 1/\sqrt{N}$, a significant reduction in noise from cosmic ray particles is required.

Fig.~\ref{proton} shows the cosmic ray background (secondary particles, such as muons, electrons, and hadrons with energies above 100 MeV) as a function of the distance from the shower axis. 
Our simulation set consists of $1.2 \times 10^6$ protons 
incident upon the atmosphere isotropically with zenith angle from $0^0$ to $60^0$. 
and energies above 10 GeV.
We assume a power law differential energy spectrum $E^{-2.8}$.
Here we consider two physical locations.
The top panel in Fig.~\ref{proton} is for New-Tupi location (sea level) and the bottom panel shows the simulation results for the HAWC experiment location (4100 m a.s.l.)\cite{abey12}. Also, the geomagnetic conditions of both locations have been taken into account.

\begin{figure}
\vspace*{0.0cm}
\hspace*{0.0cm}
\centering
\includegraphics[width=10.0cm]{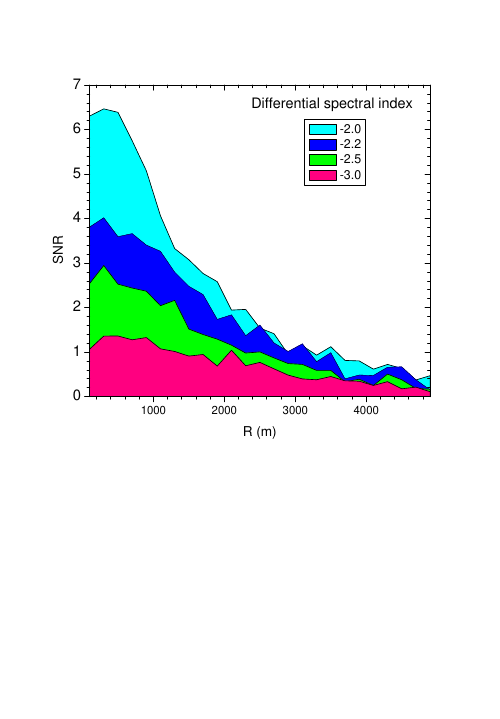}
\vspace*{-5.5cm}
\caption
{
Distribution of the signal-to-noise ratio as a function of the lateral distance to the detector. The SNR was obtained according to Eq.~\ref{eq:1}.  The Monte Carlo simulation consists of $1.2 \times 10^6$ primary photons incident upon the atmosphere with an angle of $45^0$ relative to the zenith of New-Tupi location. We have considered photons with energies above 10 GeV and a power law differential spectrum ($E_p^{-\beta}$) and for four different values of the index $\beta$.
}
\label{snr}
\end{figure}

\subsection{Estimation of the signal-to-noise ratio}
\label{subsec:snr2}

Noise-free data can not be achieved in experiment. 
However, a Monte Carlo calculations with some real input conditions can generate a free-noise signal and a pure noise signal. 
As a result, one can obtain the signal-to-noise ratio.

If the Monte Carlo calculations take into account the experimental setup properly, then the significance of the simulated signal is expected to be close to the one observed in the raw data. 
Hence, the validation of the Monte Carlo calculations can be achieved by a direct comparison of the results. 

When the signal is a transient one (e. g., a short duration peak (excess) in the time profile of the particle counting rate), then the signal-to-noise ratio (SNR) can be  defined as
\begin{equation}
SNR=\frac{N_S/\epsilon \times \Delta t}{\sqrt{N_N/\epsilon \times \Delta t}},
\label{eq:1}
\end{equation}
where  $N_S$ and $N_N$ are the numbers of  secondary particles from gamma rays (signal)
and from cosmic rays (background) in the detector during a  $\Delta t$ and that is the duration of the excess.
The $\epsilon=\epsilon_e \; \epsilon_f$ factor take into account the detection efficience ($\epsilon_e \sim 0.95$) and the fiducial volume of the detector is 90\% of the volume total ($\epsilon_f=0.9$).
 
Fig.~\ref{snr} shows the SNR obtained via Monte Carlo simulation under the following hypothesis: (a) The background particles (muons, electrons, and
hadrons) with energies above 100 MeV at sea level  using $1.2\times 10^6$ primary protons with energies above 10 GeV and a power law
differential spectrum like $E_p^{-2.8}$,
incident upon the atmosphere isotropically with 
zenith angle from $0^0$ to $60^0$.
(b) The signal particles (muons, electrons, and
hadrons) with energies above 100 MeV at sea level using $1.2\times 10^6$ primary photons with energies above 10 GeV and a power law
differential spectrum like $E_p^{-\beta}$, to four different spectral indices, $\beta=-2.0$, $-2.2$, $2.5$, and $-3.0$  reaching the top of the atmosphere during 2.2 s, and with an inclination of 45 degrees relative to the zenith.
In both cases, we have taking into account, the limitations of New-Tupi detector, such as the detection efficiency and the size of the active volume (fiducial volume).
The simulation can reproduce the SNR consistent with New-Tupi particle excess associated to the GRB 160709A only to a spectral index close to -2.2, it means
a non-steep spectrum.

 \begin{figure}
\vspace*{0.0cm}
\hspace*{0.0cm}
\centering
\includegraphics[width=12.0cm]{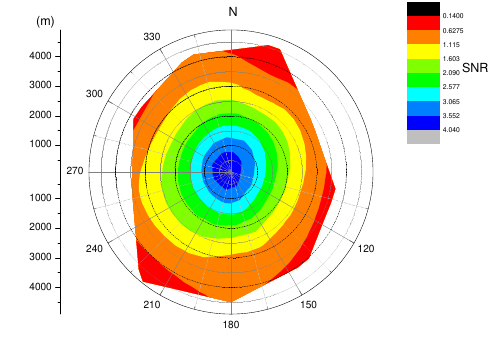}
\vspace*{-0.5cm}
\caption
{ SNR contour map for the case of a spectral index $\beta =-2.2$ and under the same simulation inputs used in the previous figure.
}
\label{snr_3}
\end{figure}

On the other hand, a steeper power law function for the gamma rays energy spectrum, let's say,  with a spectral index smaller than $-2.2$, the SNR turn very small and decreases as the spectral index decreases. Figure~\ref{snr_3}
summarizes the situation. 

In addition, from Fig.~\ref{snr} and especially from Fig.~\ref{snr_3}, where the SNR contour maps are shown at sea level, allowing seeing that the effective area for detection (in scaler mode) of primary photons in the $10-100$ GeV energy band with an energy spectrum like a power-law is about $1 km^2$.

\subsection{The observed integrated time fluence}

Due to the short duration of New-Tupi signal being in association with the prompt emission of GRB 160709A, the observed integrate time fluence can be estimated in two steps:
Firstly, we obtain of the observed signal flux 
through the relation $Flux = Rate/(G \;T_{90}\; \varepsilon)$, where Rate($=38$) is the observed counting rate of the signal (background suppressed), during the $T_{90}(=2.2 \; s)$ interval, $\epsilon (=0.95$) is the detector efficiency and  G($=35,341.9 \; cm^2 sr$) is the geometric factor of New-Tupi detector (scaler mode). Secondly, the integrated time 
fluence is calculated as
\begin{equation}
F=Flux \; T_{90}\; \Omega \; E_{thr},
\end{equation}
where $\Omega(=\pi)$ is 
the angular aperture (escaler mode). Thus, the observed integrated time fluence is 
$F=(7.4\pm 1.8)\times 10^{-4}\; GeV/cm^2$ or
$F = (1.2 \pm 0.3)\times 10^{-6} \;erg/cm^2$ in the GeV energy region.


\section{Discussion and Conclusions}
\label{sec:conclusion}

The main aim of New-Tupi detector located in the central region of the SAA is the study of solar transient events \cite{augu17}. 
It has a duty cycle (the proportion of time during which it is operated) $\sim 80\%$. 
Besides  the solar transient events study, a semi-automatic system was implemented in order to carry out a systematic search for GRB counterparts at ground in association with the satellite observations.

We have reported a search from June 2014 to February 2017, in this period Fermi LAT detected 46 GRBs with photon energies above 20 MeV. Seven of these GRBs has their coordinates within the FoV of New-Tupi detectors and two of them, the GRB 160625B and GRB160709A have a probable counterpart observed at ground level.

Initially, we were given an emphasis only on the analysis of the GRB 160709A, because it was detected by the
New-Tupi alarm system, that looks for signals from GRBs, i.e., excesses in the counting rate, but only 100 s before and 100 s after the GRB trigger.

On the other hand, the GRB 160625B was a very long duration GRB, the signal at LAT became detectable only 181 s after trigger time at Fermi GBM and whose T90 is about 460 s. At New-Tupi there is at least a peak with a significance of $4.8\sigma$ after 438 s after the trigger in probable association with GRB160625B, because it is within the T90 duration observed by Fermi GBM. The signal at New-Tupi was  detected only after a not automatic inspection. 

In relation with the GRB 160709A a nominal statistical significance of $\sim 3.5\sigma$ was found in the time profiles of New-Tupi obtained in the scaler mode. 
The signal persists above the background with a nominal significance about $3.2\sigma$ in the 4 s binning counting rate.
However, on 9 July 2016, the number of events with nominal significance above $3.5\sigma$ was 13. We believe that the magnetosphere still was perturbed because 24 hours before, there was the registered of a minor geomagnetic storm, i.e., the 9 July 2016 day was not a quiet day. Incorporating these results in the analysis we found a effective significance of $3.0\sigma$

The probability of observing an excess by New-Tupi within the range $ T90 = 5.6 \; s $ of the prompt emission of the GRB as observed by Fermi GBM, to be a random coincidence is estimated as $P\sim 5.79 \times 10^{-5}$. In addition, the probability of an event observed by New-Tupi with a nominal confidence of $3.5\sigma$ in the time profiles not to be a false alarm is $6.28 \times 10{-3}$. 

In relation with the GRB 160625B, two peaks at New-Tupi with a significance of $4.8\sigma$ and $5.0\sigma$ at $T_0+438$ s and $T_0+558$ s respectively are in probable association with GRB160625B, detected after a not automatic inspection. In addition, there is at 
New-Tupi another peak, 415 s before
of the trigger time, with a nominal statistical significance above $6\sigma$. This event could be a precursor signal from GRB 160625B, we believe that a further analysis is required to confirm this hypothesis. 

In order to obtain the effective significance
of the signal at New-Tupi in probable association with GRB 160625B, we made only a straightforward analysis.
As the duration of GRB 160625B at Fermi LAT is above 1000 s, we have excluded from the analysis the counting rate observed at $\pm 2000$ s around the trigger time. After this procedure, there were still two remain events with a nominal significance above $4.8\sigma$, and it can be considered as the background rate of excesses of nominal significance above $4.8\sigma$. The result is an effective significance of only $\sim 2\sigma$.

On the other hand, the probability of observing an excess by New-Tupi within the range of $ T90 = 460$ s  duration of the GRB as observed by Fermi GBM, to be a random coincidence is estimated as $P=5.32\times 10^{-3}$. 

We have also performed a Monte Carlo simulation taking into account the detector location and the detection energy threshold. We found that the expected signal-to-noise ratio is compatible with the observed particle excess value in association with GRB 160709A, only if the differential index of the GRB energy spectrum be equal or higher than -2.2 (a non-steep spectrum). We estimated the primary gamma-ray spectrum and the integrated time fluence of the burst as $(1.2 \pm 0.3) \times 10^{-6} erg/cm^2$ in the GeV energy range.
This fluence is about 10 times lower than the reported by KONUS-Wind fluence in the 20 keV to 10 MeV band.


\section{Acknowledgments}

This work is supported by the National Council for Research (CNPq) of Brazil, under Grants 308025/2012-1 and 307727/2015-7.
We express our gratitude to the Goddard Space Flight Center for information and data used in this study through the GCN web page (http://gcn.gsfc.nasa.gov/).
The Fermi-LAT is the product of an international collaboration between NASA and DOE in the U.S. and many scientific institutions across France, Italy, Japan and Sweden.

\end{document}